\documentclass[11pt]{article}
\usepackage{amssymb,amsfonts,amsmath,latexsym,epsf,tikz,url}
\usepackage{graphicx}
\usepackage{float}
\usepackage{fancyvrb}

\usepackage{algorithmic}
\usepackage[linesnumbered,ruled,vlined]{algorithm2e}
\usepackage{libertine}
\usepackage{listings}

\DeclareFixedFont{\ttb}{T1}{txtt}{bx}{n}{12} 
\DeclareFixedFont{\ttm}{T1}{txtt}{m}{n}{12}

\usepackage{pythonhighlight}
\usepackage{listings}

\newtheorem{theorem}{Theorem}[section]

\newtheorem{example}[theorem]{Example}
\newtheorem{definition}[theorem]{Definition}

\begin{document}

\title{Sequence pairs related to produced graphs from a method for dividing a natural number by two}

\author{
	M. Zeynali Azim$^1$ 
	\and
Saeid Alikhani$^{2,}$\footnote{Corresponding author}
\and
   Babak Anari$^1$
}

\date{\today}

\maketitle

\begin{center}
$^1$Department of Computer Engineering, Shabestar Branch, Islamic Azad University, Shabestar, Iran

$^2$Department of Mathematics, Yazd University, 89195-741, Yazd, Iran\\
{\tt mo.zeynali@gmail.com,  alikhani@yazd.ac.ir}
\end{center}


\begin{abstract}
This paper is about producing a new kind of the pairs which we call it MS-pairs. To produce these pairs, we use an algorithm for dividing a natural number $x$ by two for two arbitrary numbers and consider their  related graphs. We present some applications of these pairs that show its  interesting properties such as  unpredictability, irreversible, aperiodicity and chaotic behavior. 
\end{abstract}

\noindent{\bf Keywords:} Diamond, DGBT, MZ-Algorithm

\medskip

\section{Introduction}
\begin{figure}
 	\begin{center} 
 		\includegraphics[width=11cm, height=4.25cm]{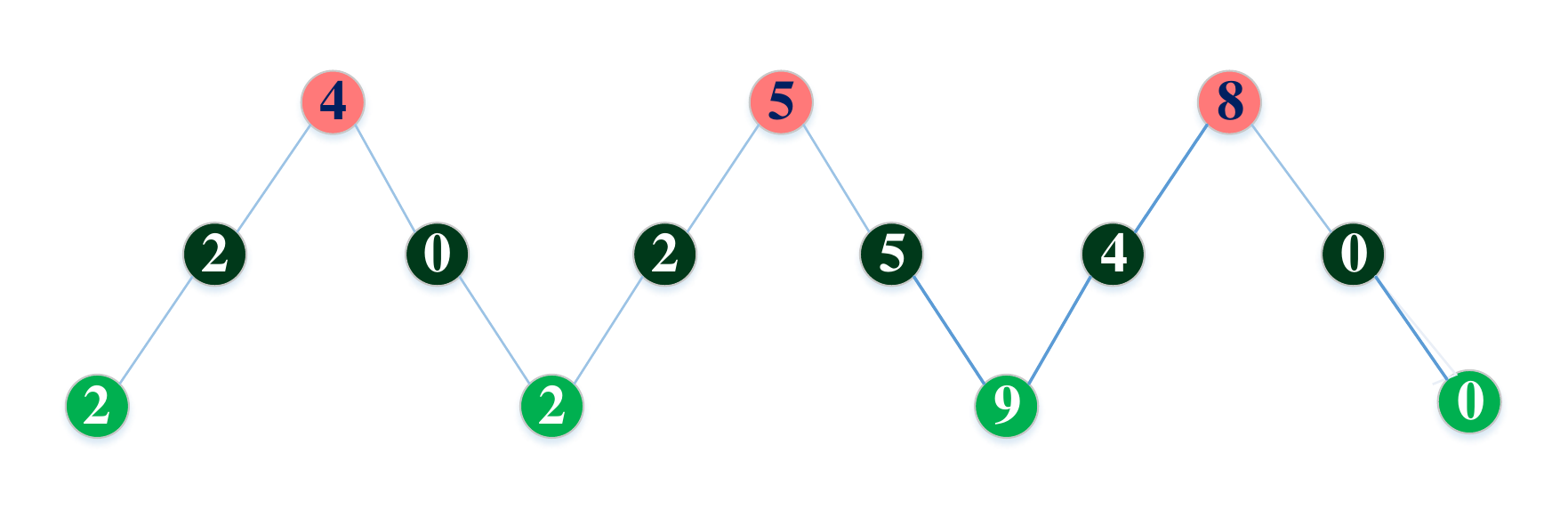}
 	\end{center} 
 	\caption{\label{fig00} The graph  $G_1(458)$.  }
 \end{figure}

\begin{figure}[!ht]
	\vspace{-.05cm}
	\begin{center} 
		\includegraphics[width=9cm, height=3.5cm]{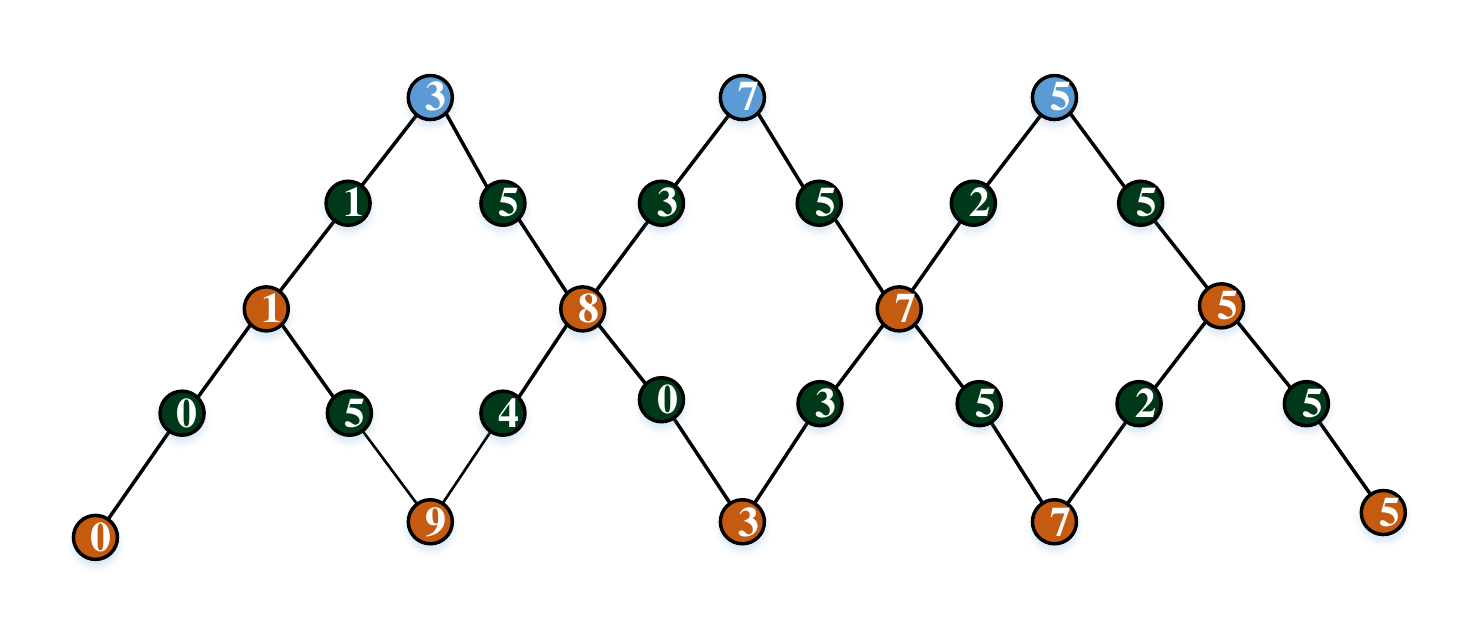}
		\caption{\label{fig02'} The graph  $G_2(375)$.}
	\end{center} 
\end{figure}

In \cite{zeynali} a new method (which is called MZ-algorithm), has presented for dividing a natural number $x$ by two and  used graphs as  models to show MZ-algorithm.
  Every digit in  a number denoted by a vertex and edges of graph draw based on MZ-algorithm. We have shown the division of the number 458 by two in Figure \ref{fig00}. This graph (Figure \ref{fig00}) which we call it division graph by two (DGBT) is a path of order $13$, i.e., $P_{13}$ (see \cite{zeynali}).    
  Applying  $k$-times of the MZ-method for the number $x$, creates  a  graph with unique structure that is denoted by $G_k(x)$ and is called DGBT. It is easy to see that $G_k(n)$ is not tree for $k>1$, since the graph has cycle. See the graph $G_2(375)$ in Figure \ref{fig02'} (see \cite{zeynali}).    

\begin{figure}[H]
	
	\begin{center} \label{fig4}
		\includegraphics[width=11.2cm, height=6cm]{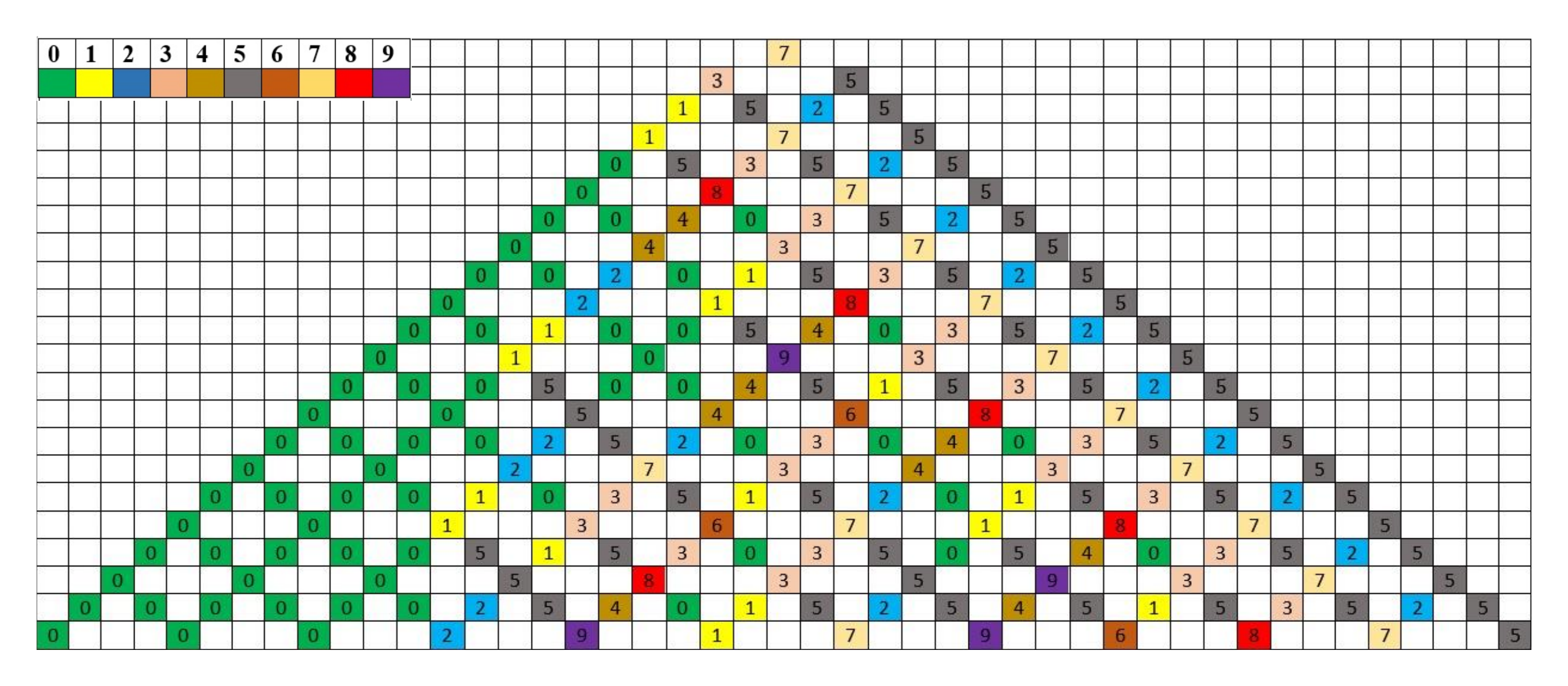}
	\end{center} 
	\caption{\label{fig04} Bitmap model for the graph  $G_{11}(7)$ }
\end{figure}

Since DGBT is an infinite graph, sometimes is better to show some of DGBT in bitmap model. In bitmap model each numbers in ${0,1,2,...,9}$ represented by unique color. See Figure \ref{fig04}.

\begin{figure}[!ht]
	\centering
	\includegraphics[width=5cm]{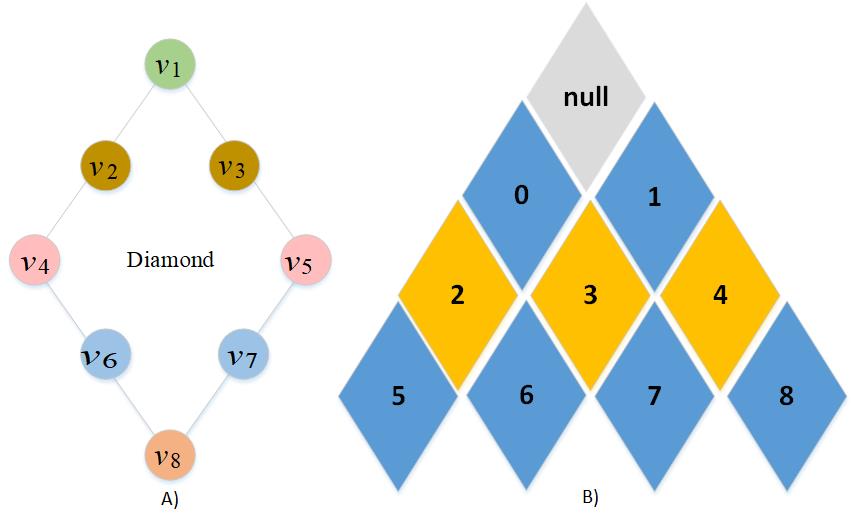}
	\caption{ A) Diamond structure. B) Nine diamonds related to level $1$ to $11$ in DGBT.}\label{vfig1}
\end{figure}

It is easy to see that the number of cycles $C_8$ in the graph $G_k(x)$ is $\frac{k-1}{2}(2d+k-2)$. We show the figures of these cycles $C_8$ in $G_k(x)$ similar to diamond.  After finding all diamonds in the graph $G_k(x)$, we label them by number $0,1,2,...$  and write these label inside  diamond (Figure \ref{vfig1}).

We consider these diamond structures and use it to produce a sequence of pairs, that we call these pairs MS-pairs.
In Section 2, we introduce MS-pairs and investigate some of its properties. In Section 3, we present some of applications and behavior  of these pairs in cryptography and chaos theory.  

\section{MS-pairs Generation}
We start this section by introducing  MS-pairs. First we state the following definition: 
\begin{definition} 
	Consider the diamond $d_i$ from the graph $G_k(x_1)$, and the diamond $d_j$ from the graph $G_k(x_2)$. We say that these two diamonds are equal, if all values of vertices in the diamond $d_i$ are equal to  all values of vertices in the diamond $d_j$. Note that we compare the value of a vertex in the diamond $d_i$ with the value of the same vertex  in the diamond $d_j$. 
\end{definition}

The following statement gives the definition of MS-pair related to graph $G_k(x_1)$ and $G_k(x_2)$: 

\begin{definition}
	The MS-pair related to graphs  $G_k(x_1)$ and $G_k(x_2)$ denoted by $(i,j)$ and the first and the second components of this pair obtain as follows. If the diamond $d_i$ from the graph $G_k(x_1)$ is equal to diamond $d_j$ from the graph $G_k(x_2)$, then we consider the label of these two diamonds as the first and the second component of a pair, i.e., $(i,j)$. 
\end{definition}

Observed that by considering the graphs $G_k(x_1)$ and $G_k(x_2)$ and applying MZ-algorithm $k$-times (for large enough number $k$) we can produces sequences of MS-pairs. These pairs have interesting properties such as  unpredictability, irreversible, aperiodicity and chaotic behavior and so they are useful and applicable in  cryptography, chaos theory, random number generation, stegnaography, password hasshing, and unique identifier generators. 
Note that if we have all of the MS-pairs that generated from $G_k(x)$ and $G_k(y)$, we cannot predict or determine the root numbers of graphs, i.e., $x$ and $y$.

\begin{figure}[!ht]
	\vspace{-.05cm}
	\begin{center} 
		\includegraphics[width=12.73cm, height=6.5cm]{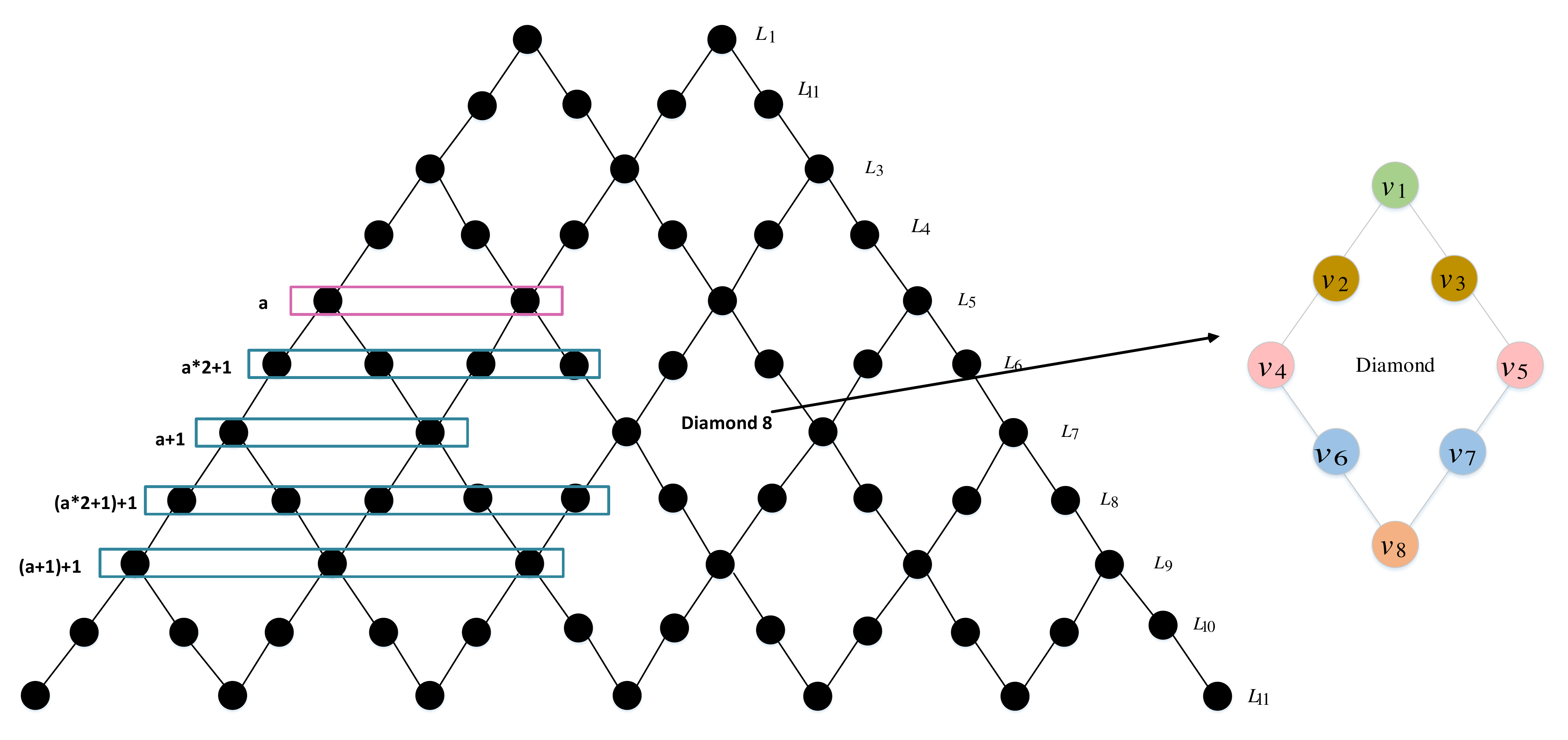}
		\caption{\label{fig2'} The graph  $G_5(n)$.  
		}
	\end{center} 
\end{figure}

The following are some of the properties of diamonds.  We consider and use Figure \ref{fig2'} for convenience. 
According this figure, the layers denoted by  $1$ to $n$.
The vertices  of a diamond is labeled with $v_1, v_2, ..., v_8$. As observe that  the first vertex i.e., the vertex $v_1$ of all diamonds are in the odd levels. Also 
we need at least five layers in DGBT to have diamond structure.  

Let use the notation $dg$ for the  number of nodes in the root of DGBT (which is the number of digits in the  number that used for constructing DGBT).

With these notation, we state the following easy theorem: 

\begin{theorem}  \label{onc}
	Let $L$ be the level of the graph DGBT. 
	\item [(i)]  If $L$ is odd, then the number of nodes in that level is 	$onc=\lfloor \frac{L-1}{2} \rfloor+dg.$
	\item [(ii)]  If $L$ is event, then the number of nodes in that level is $enc=2\left \lfloor \frac{L-1}{2}\right \rfloor+dg .$
	\item [(iii)]  The total number of nodes from the level $L_1$ to the level $L_n$ is
	$$tn=\sum_{L=1}^{n}(\left \lfloor \frac{L-1}{2}+dg \right  \rfloor)((L+1)_2+1),$$
	where the notation $(x)_2$ is the reminder of the division of $x$ by two.  
\end{theorem}


We end this section by the following  properties:  

\begin{theorem} 
	\begin{enumerate}
		\item [(i)]
		If $d$ is the number of diamonds in the DGBT graph $G_l(n)$, then  
		$$d={l+1\choose 2}-{n\choose 2}.$$
		
		\item[(ii)] If $G_k(x)$ is  an infinite DGBT, then the number of its diamonds is infinite.  
		
		\item[(ii)] 
		MS-pairs are unique. In other words, there is no two numbers $x$ and $y$ such that 
		MS-pairs produced by $G_k(x)$ and $G_k(y)$ are exactly the same.  
		
	\end{enumerate}
\end{theorem}

\section{Some applications of the MS-pairs }

In this section we state a behavior and some applications  of the MS-pairs. 
\subsection{A behavior of MS-pairs } 
In chaos theory, the butterfly effect is the sensitive dependence on initial conditions in which a small change in one state of a deterministic nonlinear system can result in large differences in a later state \cite{5}. In this subsection we observe that  a small change in the root number of DGBT, causes very large changes in the correspond  MS-pairs.

\begin{figure}[!ht]
	\centering
	\includegraphics[width=9cm]{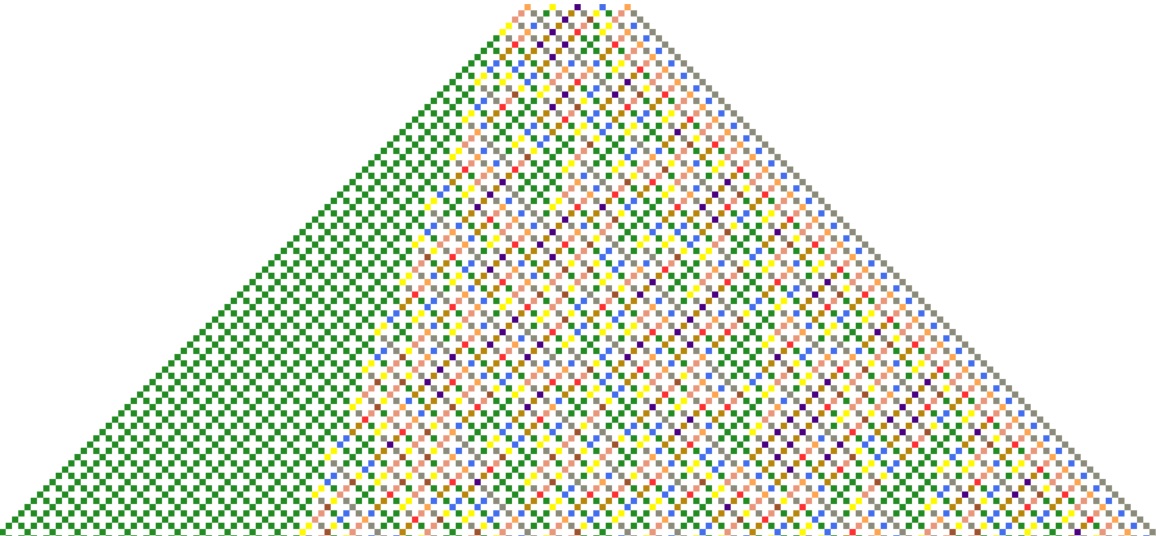}
	\caption{Bitmap model for DGBT of time 7:19:27.}\label{vfig2}
\end{figure}

Consider two graphs $G_k(x)$ and $G_k(y)$, where $x$ is a number that we can consider it as a time (which is a number with at most six digit, for example the time 7:19:27 consider as number 71927) and $y$ is an arbitrary constant number.  We have shown the DGBT for the time 7:19:27 in Figure \ref{vfig2}.

\begin{figure}[H]
	\hspace{4cm}  
		\includegraphics[width=19cm, height=13cm]{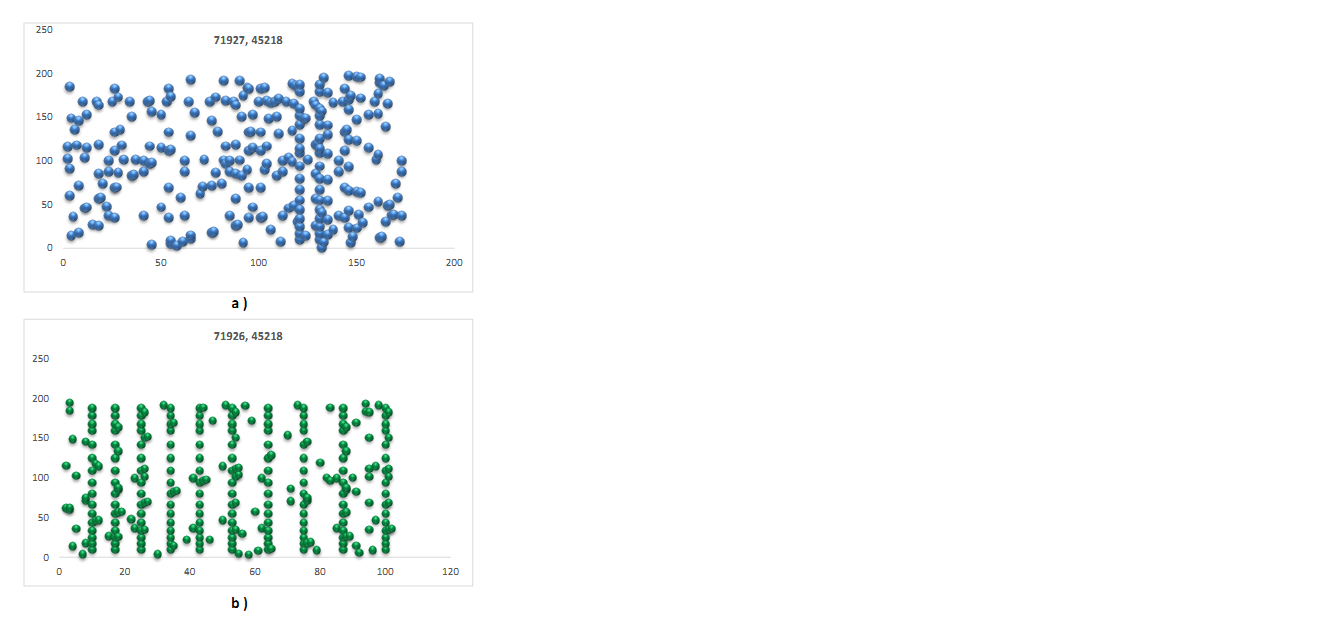}
	\vspace*{-7mm} 
	\caption{a) MS-pairs produced using $x=7:19:27$, and $y =45218$.  
		 b) Related to MS-pair that produced by  $x=7:19:26$, and $y =45218$.  
		 }\label{4}
\end{figure}

\begin{example} \label{ex1}
	Suppose that $x=7:19:27$ and $y =45218$. By Theorem \ref{onc}(iii), we have $11438$ pairs until level $100$, i.e., in 
	$G_{100}(x)$ and $G_{100}(y)$. 
	We bring up some of these MS-pairs in the following:
	
	$\big\{(2, 103), (2, 116), (2, 212), (2, 238), (3, 60), (3, 91), (3, 185), (3, 239), (4, 14), (4, 149),\\ (4, 331), (4, 360), (5, 36),
	 (5, 257), (6, 136), (6, 248), (6, 329), (7, 118), (7, 233), (7, 260), (8, 18),\\ (8, 72), (8, 146), (8, 213), (8, 261), (8, 361), (8, 374), (10, 168), 
	 (11, 46), (11, 104), (11, 204), \\(12, 47), (12, 115), (12, 153), (12, 281),...\big\}.$
	
	Now if we change time $x$ to $x_1=7:19:26$ we have $11136$ pairs in $G_{100}(x_1)$ and $G_{100}(y)$. The following are some of these MS-pairs:
	
	$\{(2, 62), (2, 116), (2, 356), (2, 390), (3, 60), (3, 63), (3, 185), (3, 195), (3, 306), (3, 391),\\ (4, 14), (4, 149), (5, 36), (5, 103), (5, 385), (6, 248),
	 (7, 4), (8, 18), (8, 72), (8, 75), (8, 146),\\ (8, 331), (8, 374), (9, 332), (10, 10), (10, 17), (10, 25), (10, 34), (10, 44), (10, 55), (10, 67), \\(10, 80),
	  (10, 94), (10, 109), (10, 125), (10, 142), (10, 60),...\}$.
\end{example} 
We compare these two sequences of pairs. Only the pairs $$\big\{(2, 116), (3, 60), (4, 14), (5, 36), (6, 248), (8, 18), (8, 146)\big\}$$
 are equal from these two sequence of MS-pairs.
As we see,  a small change in the root number, causes very large changes in the correspond  MS-pairs. This phenomenon is referred to butterfly effect in chaos theory \cite{1}.  
In order to evaluate and comparing this properties, we draw scatter plot for two MS-pairs generated in Example \ref{ex1}. Please see Figure \ref{4}.
In Figure \ref{4}, there are two figures. 
For generating Figure \ref{4}a), we use the pairs that produced by $x=7:19:27$, and $y =45218$ that displayed with blue color. 
Figure \ref{4}b) is related to MS-pair that produced by  $x=7:19:26$, and $y =45218$ displayed with green color. Observe that the two forms in these two figures are very different.

\subsubsection{Non-repudiation system by MS-pairs}
Non-repudiation is the assurance that someone cannot deny the validity of something. Non-repudiation is a legal concept that is widely used in information security and refers to a service, which provides proof of the origin of data and the integrity of the data. In other words, non-repudiation makes it very difficult to successfully deny who/where a message came from as well as the authenticity and integrity of that message.

Digital signatures (combined with other measures) can offer non-repudiation when it comes to online transactions, where it is crucial to ensure that a party to a contract or a communication cannot deny the authenticity of their signature on a document or sending the communication in the first place \cite{4}.

Here, using MS-pairs we prove the claim   of one of the two parties (users or systems). 
As an example, we consider  two persons $A, B$, where $A$ has ID-number $15963$ and $B$ has  ID-number $15964$. At the time $t=7:19:27$ person 
$A$ have signed the contract $d$. We  generate MS-pairs with  $t=7:19:27$,  ID-number=$15963$ and write the produced MS-pairs on the contract $d$.
Suppose that after signing the contract $d$ by $A$,  two person $A, B$ both claim that in the time $t$, signed or not-signed the contract $d$. 
To judge  their claims, system or the third person $C$  get their ID-numbers
and produce MS-pairs for these ID-numbers and the time $t$ on the contract  $d$.  If the produced MS-pairs are equal to the MS-pairs on the contract,  then the claim of that person is true.
For example, the MS-Pairs generated for the person $A$ is:\\ 
$MS-Pairs(7:19:27, 15963)=[(4, 61), (5, 73), (6, 1), (7, 2), (7, 29), (8, 3), (9, 73),\\ (10, 10), (10, 17), (10, 25), (10, 34), (10, 44), (10, 55), (10, 67), (10, 80), (10, 94),(13, 7),\\ (13, 13), (13, 20), (14, 8), (16, 24), (16, 43), (16, 66), (16, 93), (17, 10), (17, 17), (17, 25),\\ (17, 34), (17, 44), (17, 55), (17, 67),...]$ and \\
The MS-Pairs generated for the person $B$ is:\\ 
$MS-Pairs(7:19:27, 15964)=[(1, 62), (2, 75), (4, 61), (5, 73), (6, 1), (6, 74), (7, 2),\\(7, 29), (7, 88), (9, 73), (10, 23), (10, 32), (10, 42), (10, 53), (10, 65), (10, 78), (10, 92),\\ (12, 87), (13, 7), (13, 13), (13, 20), (16, 31), (16, 52), (16, 77), (17, 23), (17, 32), (17, 42),\\ (17, 53), (17, 65), (17, 78), (17, 92),...]$\\

The following algorithm, gives the non-reproduction steps of the proposed algorithm in more detail.

\medskip

\IncMargin{1em} \begin{algorithm} [H]\label{111}
	\SetKwData{Left}{left}\SetKwData{This}{this}\SetKwData{Up}{up} \SetKwFunction{Union}{Union}
	\SetKwFunction{FindCompress}{FindCompress} 
	\SetKwInOut{Input}{input}\SetKwInOut{Output}{output}
	\Input{$ID_A, ID_B, time\ t,\ contract\ with\ d\ ID$ }
	\Output{$MS-pairs, claim~true$} 
	$get(ID_A, t)$\\
	$d\leftarrow\ {produce\_pairs(ID_A, t)} $  \;
	$ {get(ID_A, ID_B, t)} $  \;
	$claim_A\leftarrow\ {produce\_pairs(ID_A, t)} $  \;
	$claim_B\leftarrow\ {produce\_pairs(ID_B, t)} $  \;
	\If { $ d==claim_A$}{
		$print(``A\ claim\ is\ true")$
	}
	\If { $ d==claim_B$}{
		$print(``B\ claim\ is\ true")$
	}
	
	\caption{Non-repudiation System Algorithm}\label{non-reproduction Algorithm} \end{algorithm}\DecMargin{1em}\


\subsection{Applications to Steganography and Cryptography}

\medskip
Stream cipher is an important branch of symmetric cryptosystems, which takes obvious advantages
in speed and scale of hardware implementation. It is suitable for using in the cases of massive data
transfer or resource constraints, and has always been a hot and central research topic in cryptography.  
A word-oriented stream cipher usually works on and outputs words of certain size, like 32, 16, 8 bits \cite{4}.
We will use MS-pairs for word-stream-cipher cryptographic application.  We present a simple algorithms to show that how MS-pairs could to encrypt and decrypt data or message. 
Word size in this algorithm is 8 bits. 

\IncMargin{1em} \begin{algorithm} [H]\label{111}
	\SetKwData{Left}{left}\SetKwData{This}{this}\SetKwData{Up}{up} \SetKwFunction{Union}{Union}
	\SetKwFunction{FindCompress}{FindCompress} 
	\SetKwInOut{Input}{input}\SetKwInOut{Output}{output}
	$Sender\_Select:\ x_1, diamond\_number$\\
	$Sender\_Send:\ x_1, diamond\_number\to\ reciver$
	
	$Reciver\_Select:\ x_2$\\
	$MS-pairs\leftarrow Reciver\_Compute\_algorithm1(x_2,x_1)$\\

	\While{$p\ in\ MS-pairs$}{
		$e_i\leftarrow ((p_a+p_b)\ mod\ Chara\_number)$\;    
		$i=i+1$ \\  
		$en_i\leftarrow (|e_i-char\_code|)$\;  
		
	}
	
	$Output\leftarrow\ {en} $  \;   
	\caption{MS-Pairs encryption}\label{Pairs Generation Algorithm} \end{algorithm}\DecMargin{1em}\

Decryption algorithm is as follows:

\IncMargin{1em} \begin{algorithm} [H]\label{111}
	\SetKwData{Left}{left}\SetKwData{This}{this}\SetKwData{Up}{up} \SetKwFunction{Union}{Union}
	\SetKwFunction{FindCompress}{FindCompress} 
	\SetKwInOut{Input}{input}\SetKwInOut{Output}{output}
	$Sender\_recived:\ x_2, en$\\
	$MS-pairs\leftarrow Reciver\_Compute\_algorithm1(x_2,x_1)$\\	
	
	\While{$p\ in\ MS-pairs$}{
		$d_i\leftarrow ((p_a+p_b)\ mod\ Chara\_number)$\;    
		$i=i+1$ \\  
		$de_i\leftarrow (|d_i-char\_code|)$\;  
		
	}
	
	$Output\leftarrow\ {de} $  \;   
	\caption{MS-Pairs decryption}\label{Pairs Generation Algorithm} \end{algorithm}\DecMargin{1em}\

For example we encrypt  and decrypt the plain-text ``Graph" with arbitrary $x_1=71927,x_2=45218$ and $diamond\_number=13$. 
First we produce MS-Pairs for $x_1, x_2$ then, we take the number of characters in the message from the MS-Pairs.
According this algorithm since the plain-text has five characters so we can 
select only five pairs.
Generated pairs are:
$$[(7, 118), (7, 233), (7, 260), (8, 18), (8, 72)].\\$$
We consider arbitrary character number, say $256$. According the encryption algorithm we have:

 $ e_1=(7+118)=125 ~(mode\ 256)$,
 
  $ e_2=(7+233)=240 ~(mode\ 256)$,
  
    $e_3=(7+260)=11 ~ (mode\ 256)$,
    
     $e_4=(8+18)=26 ~ (mode\ 256)$,
     
      $ e_5=(8+72)= 80~(mode\ 256)$. 
      
      Then we discover the code of characters in ``Graph". We have: 
      
      $char\_code\leftarrow\{71, 114, 97, 112, 104\}$ and 
      
$en=\{|71-125|, |114-240|,|97-11|, |112-26|, |104-80|\}$. 

Finally,  the encryption of "Graph" will be: $en=\{54, 126, 86, 86, 24\}$. As we know that the decryption algorithm is reverse of encryption. So we can reach from $en=\{54, 126, 86, 86, 24\}$ to ``Graph" by using decryption algorithm.
For decryption we can do the reverse operation on encrypted message. According Algorithm 5,  we have
$$en=\{71, 114, 97, 112, 104\}$$




\end{document}